# The choice of viscous or viscoelastic models affects attenuation and velocity determination in simplified skull-mimicking digital phantoms

Authors

Samuel Clinard[1,2*], Taylor Webb[2], Henrik Odéen[2], Dennis L. Parker[2,1], Douglas A. Christensen[1,3]

1. Department of Biomedical Engineering, University of Utah, Salt Lake City, Utah, USA
2. Department of Radiology and Imaging Sciences, University of Utah, Salt Lake City, USA
3. Department of Electrical and Computer Engineering, University of Utah, Salt Lake City, U

*Corresponding Author

Sam.Clinard@utah.edu
540-326-5601

729 Arapeen Dr.
Salt Lake City, Utah 84108

Abstract:

**Objective:** Simulation-guided transcranial focused ultrasound therapies rely on accurately estimating the skull's acoustic properties with pretreatment imaging. The typical imaging resolution (0.5 mm isotropic) is insufficient to resolve the bone microstructure, and as a result, the acoustic properties are underdetermined. Consequently, the determination of acoustic properties is sensitive to methodology. Here, we investigate how viscous or viscoelastic models predict variations in attenuation and phase velocity due to microstructure.

**Approach:** Using viscous and viscoelastic k-Wave implementations, we simulated the transmission of a broadband 625 kHz tone burst ( -6 dB bandwidth: 250 kHz–1 MHz) through skull-mimicking digital phantoms. The phantoms consisted of spherical marrow pores (0.1 mm–1.0 mm diameter) randomly placed within a solid cortical background (2.5%–90% porosity). Virtual sensors measured the attenuation and phase velocity by the time-distance matrix approach.

**Results:** Both models predict that attenuation increases with increasing pore size for a fixed porosity, but the strength of the relationship and its dependence on porosity depend on the simulation technique applied. The viscoelastic model generally predicts that the attenuation peaks at larger porosity values than those predicted by the viscous model. For example, for 1.0 mm pore phantoms, the viscous attenuation peak (1.98 Np/cm) occurs at 20% porosity, while the viscoelastic peak (2.98 Np/cm) occurs at 70%. In general, the phase velocity decreases for larger pore phantoms for both techniques, but the phase velocity/porosity relationship predicted by the viscoelastic model is less sensitive to pore size compared to the viscous model.

**Significance:** Viscous and viscoelastic models predict different attenuation and phase velocity behavior for these idealized bone models. While both approaches predict that properties of bone microstructure have a large impact on acoustic attenuation within the skull, the viscous model indicates that bone microstructure impacts phase velocity within the skull, whereas the viscoelastic model suggests that phase velocity is less sensitive to microstructure. This study highlights the acoustic model's role in predicting transcranial propagation and the need to identify which approach most accurately reflects the physics of the human skull.

## Introduction

Transcranial focused ultrasound (tcFUS) is a promising noninvasive modality for treating neurological disorders. Ultrasound can be focused within deep brain regions with various effects on brain tissue, including thermal ablation, neuromodulation, and opening of the blood-brain barrier (BBB). Clinically, high-intensity focused ultrasound (HIFU) can treat Essential Tremor and Parkinson's Tremor with thermal ablation (Elias *et al* 2016, Krishna *et al* 2023), while low-intensity focused ultrasound (LIFU) is under investigation for several neurological disorders including depression and chronic pain, and, when used with microbubbles to facilitate BBB opening, for drug delivery in treating brain tumors and neurodegenerative diseases (Schachtner *et al* 2025, Shi and Wu 2025, Karmur *et al* 2020).

A major challenge in tcFUS is compensating for the heterogeneous patient-specific skull. The skull bone has significantly different acoustic properties from soft tissue, leading to aberration and attenuation of the transmitted ultrasound beams. Without correction, the skull degrades the focal quality, potentially reducing treatment efficacy and safety. HIFU treatments utilize pretreatment simulation guidance for correction, treatment monitoring, and patient response to deliver ablative therapies precisely (Quadri *et al* 2018). Magnetic resonance temperature imaging (MRTI) is essential for these therapies as it confirms the focal location and provides monitoring of the ablation utilizing a thermal dose (Odéen and Parker 2018). Notably, LIFU treatments cannot be monitored with MRTI as the temperature rise is typically less than 2 °C (Aubry *et al* 2025). As such, alternative methods to ensure precise ultrasound delivery for safe and effective tcFUS interventions are needed.

Improved ultrasound simulation guidance methods may improve the precision in transcranial ultrasound delivery and ultimately enable transcranial focusing for LIFU. Simulations are currently limited by uncertainty in the skull's acoustic properties, including attenuation and phase velocity. (Angla *et al* 2023). These acoustic parameters are typically derived from pre-treatment computed tomography (CT); however, there is no consensus on which of several reported relationships should be used clinically (Leung *et al* 2019). Further, our previous work demonstrated that no unique relationship exists between attenuation, phase velocity, and CT Hounsfield Units (Clinard *et al* 2025b, 2025a) at typical clinical CT resolution (~0.5 mm isotropic). This resolution cannot reveal the skull's microstructure and its effect on ultrasound propagation. Because acoustic properties are underdetermined by CT, any CT-based parameter estimation method will likely be sensitive to the employed methodology.

Simulations can employ both viscous and viscoelastic acoustic models of skull bone. Parameter-estimation studies have primarily used viscous fluid models (Aubry *et al* 2003, McDannold *et al* 2019, Leung *et al* 2021, Vyas *et al* 2016), while a few used viscoelastic solid models (Pichardo *et al* 2017, Pinton *et al* 2011). Viscous models are typically less computationally complex, requiring simpler calculations and less memory (Vyas and Christensen 2012, Treeby and Cox 2010a). Viscous fluids only support longitudinal waves while viscoelastic solids support both longitudinal and shear waves. Skull bone can support both wave modes, but mode conversion is limited for near-normal incident wave interfaces (White *et al* 2006, Hayner and Hynynen 2001).

Studies comparing viscous and viscoelastic models of microstructural interactions have been limited. Specifically, the scattering behavior of the microstructure strongly determines both the attenuation and phase velocity (Pinton *et al* 2011, Haïat and Naili 2011). Classic scattering physics predicts that even in the simple case of a plane wave incident on a spherical pore, viscous and viscoelastic models predict different scattering magnitudes and directions (Faran 1951, Ávila-Carrera and Sánchez-Sesma 2006). The complexity of scattering increases within heterogeneous bone, and viscoelastic interactions with bone microstructure will affect attenuation and velocity dispersion (Mézière *et al* 2014, Haire and Langton 1999). Importantly, the extent to which the model choice affects acoustic parameter estimation of skull bone has not yet been determined.

This study employs viscous and viscoelastic k-Wave models. k-Wave is a well-validated simulation tool that includes full-wave pseudo-spectral time-domain solvers for both viscous and viscoelastic wave equations

(Treeby and Cox, 2010a; Aubry et al., 2022; Treeby et al.,2014b). Using these simulation methods, we determined how each model predicts attenuation and phase velocity within idealized skull bone phantoms with varying pore sizes and porosities. The results may inform future parameter estimation work, highlighting the importance of accurately modeling skull microstructure.

## Materials and Methods

We conducted full-wave ultrasound simulations of digital phantoms using both viscous fluid and viscoelastic solid k-Wave models to investigate the impact of modeling on phase velocity and attenuation. Specifically, we created a set of digital phantoms by placing pores of a single size (ranging from 0.1 mm to 1.0 mm) in a solid cortical background. Each phantom had randomly placed pores of a single specific size and targeted a specific porosity (values ranging from 2.5 to 90%). Using these phantoms, we measured the velocity and attenuation using the previously established time-distance matrix approach (Mézière *et al* 2014). Using this method, we sought to determine if viscous and viscoelastic models produce different attenuation and phase velocities within each phantom.

*Bone Phantoms*
Three-dimensional bone-mimicking phantoms (6.4 mm x 6.4 mm x 8 mm; isotropic resolution of 0.05 mm) were constructed by randomly placing spherical marrow pores of a single diameter into a solid cortical background. The first 0.6 mm consisted of a layer of solid cortical bone. The acoustic properties of the constituent materials are provided in Table 1. We note that while a small shear velocity (10 m/s) and attenuation (0.05 np/cm) have been reported for marrow (White *et al* 2006), this study approximates these values as zero to reduce computational requirements, as described in the simulation methods section. We chose four pore diameters (0.1 mm, 0.2 mm, 0.5 mm, and 1.0 mm) that are smaller than or on the order of the typical imaging resolution of CT (~0.5 mm to 1.0 mm), in order to study the effects of microstructure with features smaller than what can be resolved with clinical CT. The pores were allowed to overlap to better mimic the complex bone microstructure. Nine porosities, ranging from 2.5% to 90%, were constructed for each pore diameter, resulting in a total of 36 phantoms, each with a unique porosity/pore diameter pair. Five phantom sets with different random pore placements were generated to evaluate the dependence of the results on specific microstructures. Figure 1 illustrates the simulation geometry accompanied by 3D renderings of exemplary phantoms.

*Table 1: Acoustic Properties of Constituent Materials at 625 kHz (Hasgall et al 2022, White et al 2006)*

| Material | $c_p$ (m/s) longitudinal velocity | $α_p$ (Np/m) longitudinal attenuation | $c_s$ (m/s) shear velocity | $α_s$ (Np/m) shear attenuation | Density (kg/m³) |
|---|---|---|---|---|---|
| Red Marrow | 1450 | 7.8 | 0 | 0 | 1029 |
| Cortical Bone | 3514 | 26.7 | 1400 | 56 | 1908 |

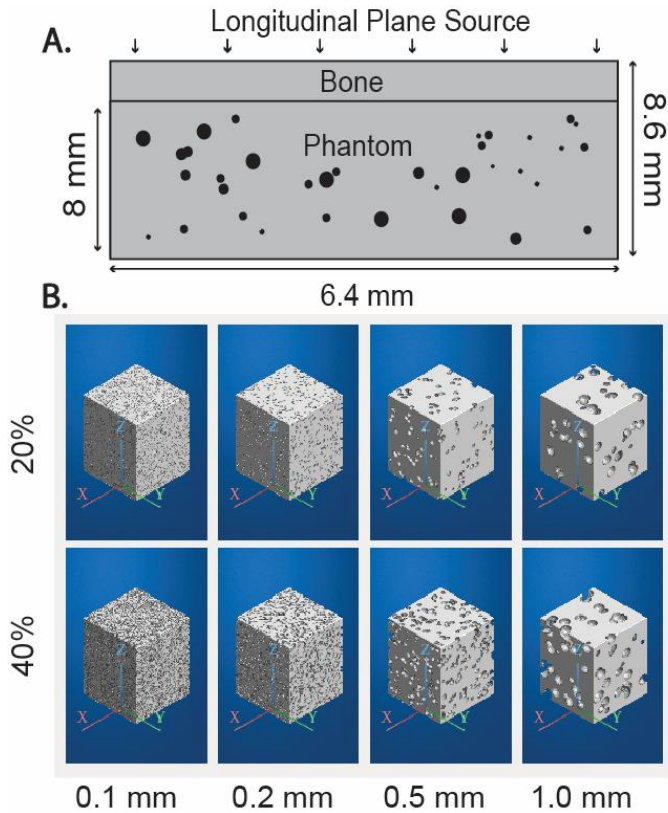

*Figure 1: (A) Simulation geometry of a longitudinal source propagating through a 0.6 mm layer of cortical bone (light gray) and an idealized phantom of cancellous bone (with dark marrow pores). The pore size is uniform; the variation in apparent size is due to the 2D slice view at a given transverse position. (B) 3D renderings of eight exemplary phantoms with two porosities (rows) and four pore sizes (columns).*

3D Simulations
All simulations were completed using k-Wave version 1.4. The time-step was defined using a Courant-Friedrichs-Lewy (CFL) Condition of 0.06. Details of spatial and temporal convergence testing are included in the supplemental material. The reference speed of sound was set to the longitudinal minimum (marrow, 1450 m/s), which reduces non-physical numerical dispersion but requires a more stringent stability criterion in the time step (Treeby et al 2016). Modeling a marrow shear velocity of 10 m/s would require a much smaller time step. As such, we approximate the marrow shear properties as 0 m/s and 0 Np/cm; therefore, the marrow does not support shear waves. The initial 0.6 mm of the grid in the slice direction were assigned solid cortical bone properties, and the digital phantoms were placed in the remaining eight mms, as illustrated in Figure 1. The starting cortical layer reduces the memory required to define the source, as it enables a spatially uniform, time-varying source, as described later. By avoiding planar water/bone interfaces, the effects of non-physical multiple reflections are reduced. A 10-voxel-thick, perfectly matched layer with 2.0 Np/voxel attenuation was added externally to the end planes to achieve an effective infinite domain. Transverse matching layers were not employed as transverse periodicity produces a plane wave when combined with a uniform source (as described in the k-Wave MATLAB examples) (Treeby *et al* 2014a).

We utilized both viscous and viscoelastic implementations of k-Wave, realized by the k-Wave functions 'kspaceFirstOrder3DG' and 'psdElastic3D'. While both functions employ a pseudo-spectral time-domain method, they solve different wave equations and require different parameters. The viscous implementation solves a linearized acoustic wave equation for a viscous, absorbing fluid that supports only longitudinal waves (Treeby and Cox 2010a). The viscoelastic implementation solves the first-order form of a viscoelastic wave equation, which supports both longitudinal and shear waves. Both implementations were assigned the same longitudinal velocities and attenuations, as well as densities (Table 1). The viscoelastic model additionally requires a shear velocity and attenuation (Table 1), from which the complex Lamé parameters ($\lambda$ and $\mu$) are derived to govern the propagation of longitudinal and shear stress fields and their mode conversion (Treeby et al 2014b).

The viscous simulations included acoustic absorption, which in k-Wave is modeled by a frequency power law that relates two loss terms to the frequency-dependent absorption, given by $\alpha(f) = \alpha_{1\ MHz} f^y$ (Treeby and Cox2010b), where $\alpha_{1\ MHz}$ is the power law pre-factor and $y$ is the frequency exponent. The exponent $y$ is related to the dispersion by the Kramers-Kronig (k-k) relationship (Treeby and Cox 2010b). We assumed the materials themselves were dispersionless to focus on dispersion due to inter-voxel path length changes. The dispersion introduced by power-law absorption can be eliminated by setting the frequency exponent to 2. This means the constituent materials are dispersionless, although dispersion may occur due to interactions within the microstructure.

The viscoelastic simulations also include acoustic absorption, which is modeled by the Kelvin-Voigt model that incorporates compressional ($\chi$) and shear ($\eta$) viscosity coefficients(Treeby et al 2014b). When the effect of the loss term is small, this model results in absorption that follows a frequency-squared power law $\alpha(f) = \alpha_{1\ MHz} f^2$ for both compressional and shear waves (Treeby et al 2014b). This relationship is intrinsic to the governing equations, and as such, the viscoelastic power law relationship cannot be adjusted. For a small loss, the viscoelastic solution exhibits no material dispersion.

All simulations employed a spatially uniform, single-cycle, broadband velocity source aligned with the propagation direction to generate a planar longitudinal wave. The source was applied on the front z-plane in bone and propagated in the z- direction. A 1 MPa pressure ($P_0$) source was achieved by setting the particle velocity ($U_0$) amplitude to $U_0 = P_0/Z_{bone}$, where $Z_{bone}$ (6.7 MRayl) is the bone's acoustic impedance. To minimize startup transient waves, we applied a Gaussian envelope to the single-cycle 625 kHz tone burst (target -6 dB bandwidth: 250 kHz–1 MHz). Because the pulse is truncated, the actual spectrum is slightly broadened. In practice, our phase velocity and attenuation calculation methods rely on relative amplitudes and phase over distance, so this small spectral discrepancy does not materially affect the results.

A virtual sensor array was defined to record the time-varying pressure along the z-axis (direction of propagation). The array consisted of rectangular sensor elements oriented transversely to the source and positioned from 1.1 mm to 6.1 mm from the model front with a 0.05 mm spacing, consistent with the grid resolution. The array bounds were empirically chosen to exclude non-physical reflections from the matching layers. The sensor array recorded the spatially averaged pressure in the viscous simulations as well as the normal stress (longitudinal pressure) in the viscoelastic simulations. The result was a pressure-time-distance matrix from which both attenuation and phase velocity can be derived, as described next.

*Attenuation and Phase Velocity Measurement*
The attenuation and phase velocity can be calculated using the previously established time-distance matrix approach (Mézière et al 2014). This involves calculating the complex pressure propagation $\tilde{s}_t(z)$ within a medium using the virtual sensors. While this approach cannot be accomplished in a physical setup, it has an advantage over the more physical transmission method as the ballistic wave is more easily isolated from reflections.

To find the phase velocity, the unwrapped phase angle in the propagation direction, $\varphi_\omega(z)$, is defined as:

$$\varphi_\omega(z) = \arg\left(\tilde{S}_\omega(z)\right) = \varphi_\omega(0) + k(\omega)z, \tag{1}$$

where $\tilde{S}_\omega(z)$ is the Fourier transform of $\tilde{s}_t(z)$ and $\varphi_\omega(z)$ is linear in z with a slope equal to $k(\omega)$. After solving for $k(\omega)$, the phase velocity can be calculated from its definition by the dispersion relation:

$$v_p(\omega) = \frac{\omega}{k(\omega)}. \tag{2}$$

The frequency-dependent attenuation can be found by the modulus of $\tilde{S}_\omega(z)$ assuming an exponential decrease in signal given by

$$\left|\tilde{S}_\omega(z)\right| = e^{-\alpha(\omega)z}. \tag{3}$$

The attenuation $\alpha(\omega)$ in Np/cm is found by taking a linear fit of the logarithm of this function.

To investigate the effect of acoustic models (viscous or viscoelastic) on the variations of attenuation and phase velocity within different bone phantom microstructures, we conducted experiments on each of our 36 phantoms. These experiments explored the relationships of $\alpha(\omega)$ and $v_p$ to porosity, pore size, and frequency.

## Results

Figure 2 graphs the relationship of attenuation and phase velocity to frequency for both viscous and viscoelastic models. Results are shown using representative 40% porosity phantoms comprised of all four pore sizes. The attenuation increases with frequency in all cases, but the values are different for the two approaches: the attenuation at the center frequency (625 kHz) for the four pore sizes is 0.22, 0.22, 0.47, and 1.66 Np/cm for viscous simulations, and 0.15, 0.27, 1.03, and 1.85 Np/cm for viscoelastic simulations. There is little variation between the two smaller pore sizes for the viscous model. In the viscoelastic model, the attenuation approximately doubles between pore sizes of 0.1 mm to 0.2 mm. The standard error across five randomizations is small for all pore sizes except the 1.0 mm pore phantoms, indicating the attenuation depends on the particular spatial pattern of pore locations for coarser microstructures.

The phase velocity mainly decreases with increasing pore size for both acoustic models. At the center frequency of 625 kHz, the phase velocity for the four pore sizes is 2622, 2413, 1992, and 1804 m/s for viscous simulations, and 2643, 2485, 2252, and 2303 m/s for viscoelastic simulations. An exception to this trend occurs when comparing the viscoelastic phase velocity between pore sizes of 0.5 mm and 1.0 mm, which shows a slight increase with increasing pore size. The variation of phase velocity with pore size is greater in the viscous case. Frequency dispersion of the phase velocity can be quantified by fitting a line from 250 kHz to 1 MHz to the curves. The dispersion (for 40% porosity) with increasing pore size is -4, -12, -104, and -130 m/s·MHz for the viscous simulations, and -19, -21, 53, and 182 m/s·MHz for the viscoelastic simulations. Note that the polarity of the dispersion depends on the simulation method and the pore size. For the 0.5 mm and 1.0 mm phantoms, the viscous dispersion is negative while the viscoelastic dispersion is positive.

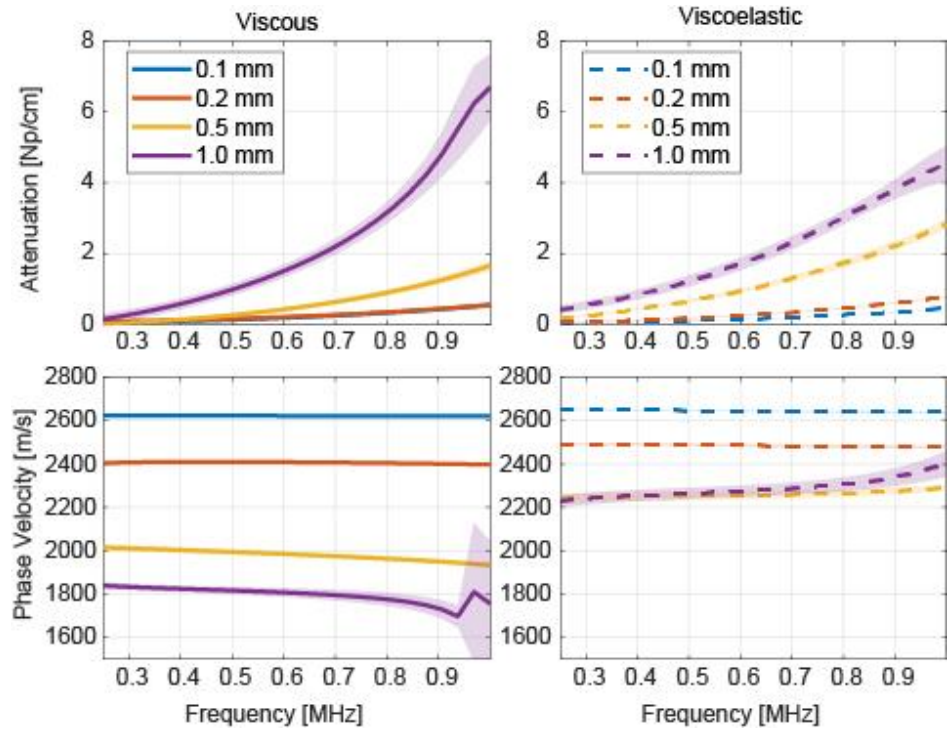

*Figure 2: Attenuation (top) and phase velocity (bottom) as a function of frequency for 40% porosity phantoms. Viscous (left) and viscoelastic (right) results are plotted for four pore sizes as indicated by the legend. The standard error over the five different randomizations is shown by the shaded areas.*

Figure 3 shows the attenuation as a function of porosity and pore size for both simulation methods at three frequencies. Consistent with the results shown in Fig. 2 at 40 % porosity, the large-pore phantoms generally attenuate more than the small-pore phantoms. For the 0.1 mm pore phantoms, there is little variation in attenuation between the two simulation methods: the attenuation is mostly linear and decreases with porosity. At 0.2 mm, the viscoelastic result exhibits a slight increase in attenuation with porosity that peaks at a porosity of 70%. For the larger pore phantoms, the attenuation first increases and then decreases with porosity; however, the peak occurs at different porosities depending on the acoustic model. For example, at 625 kHz and 0.5 mm pore size, the viscous peak is 0.49 Np/cm at 20% porosity while the viscoelastic peak is 1.38 Np/cm at 70% porosity. As evident in the results for the larger pore sizes, the acoustic model has a significant impact on relative attenuation at each porosity.

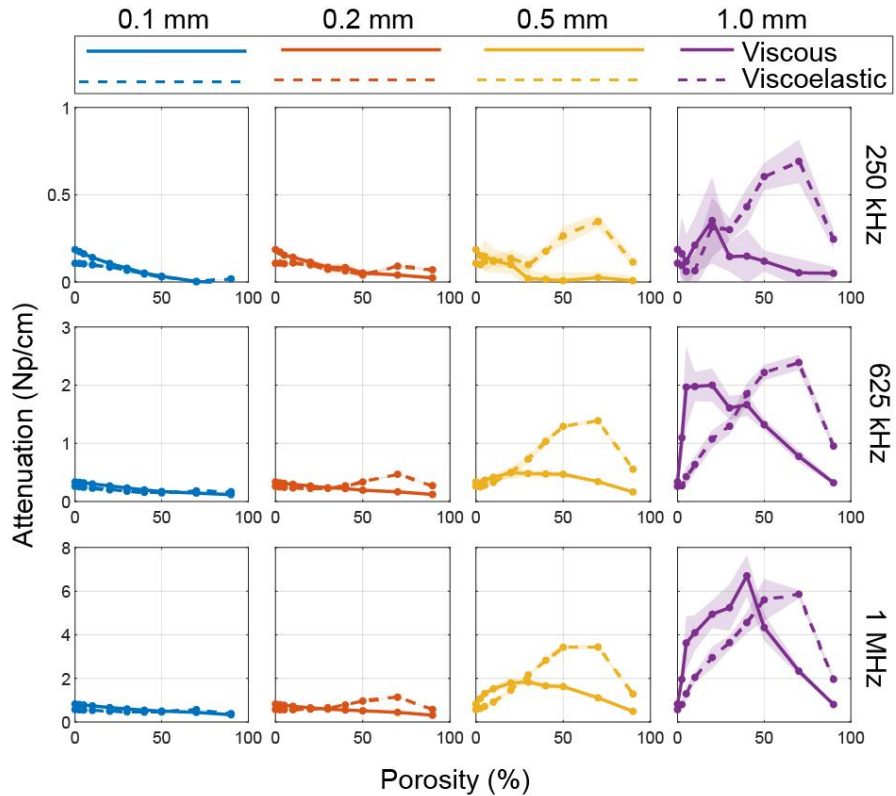

*Figure 3: Attenuation relationship to porosity at three frequencies (rows) and four pore sizes (columns). The viscous results are plotted with solid lines, while the viscoelastic results are plotted with dashed lines. The relative attenuation depends on porosity, pore size, frequency, and acoustic model. The standard error over the five different randomizations is shown by the shaded areas.*

Figure 4 shows the phase velocity as a function of porosity and pore size for both simulation methods at three frequencies. The phase velocity generally decreases with increasing pore size, and the velocity is typically higher for the viscoelastic simulations compared to the viscous case, especially for the larger pore sizes. The smaller pore phantoms exhibit a mostly decreasing, linear relationship with porosity, while the larger pore diameter phantoms have a more non-linear relationship to porosity. The phase velocity varies only slightly with frequency for both viscous and viscoelastic simulations for the smaller pore sizes, more for the larger pore sizes, consistent with Fig. 2. At 40% porosity and 625 kHz, the 0.1 mm pore phantom exhibits a viscous velocity of 2622 m/s and a viscoelastic velocity of 2643 m/s. At a pore size of 0.5 mm, the viscous velocity is 1982 m/s, and the viscoelastic velocity is 2275 m/s. The viscous model shows a larger variation with pore size compared to the viscoelastic model. The standard error is small for all cases except those involving the viscous 1.0 mm pore phantoms with porosities ranging from 20% to 50%. This may be explained by an increased sensitivity to the spatial randomization of the pores due to resonant and scattering effects.

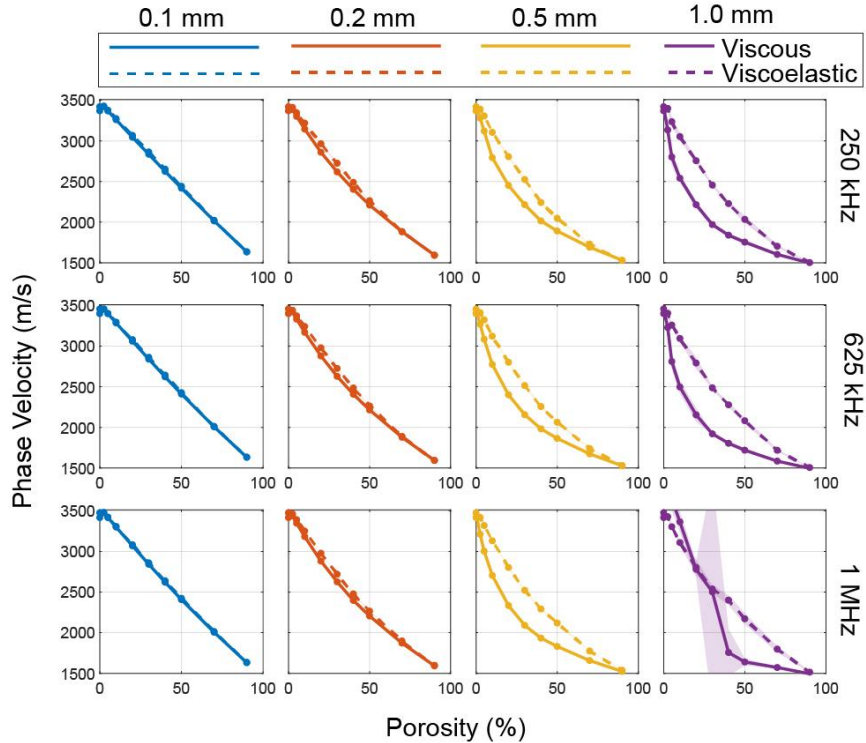

*Figure 4: Phase velocity relationship to porosity at three frequencies (rows) and four pore sizes (columns). The viscous results are plotted with solid lines while the viscoelastic results are plotted with dashed lines. The standard error over the five different randomizations is shown by the shaded areas.*

## Discussion

The choice between a viscous or viscoelastic model alters the predicted relationships between attenuation, velocity, and porosity, suggesting that parameter estimation is model-dependent. Despite these differences, both models predict that attenuation and velocity depend on pore size, confirming our previous work, which demonstrated that sub-voxel heterogeneity introduces uncertainty in CT-derived acoustic properties (Clinard *et al* 2025a, 2025b).

The two acoustic models produce different attenuation-porosity relationships (Figure 3), in part because they model scattering differently. For the simplest case of a longitudinal plane wave incident on a spherical inclusion, the analytical solutions predict different magnitudes and directions of scattering for the two acoustic models (Ávila-Carrera and Sánchez-Sesma 2006). Within more complicated heterogeneous media, it follows that multiple scattering will also behave differently. The loss is further complicated within viscoelastic models by mode conversion. The result can be characterized by an increasing loss with frequency (Figure 2); however, our results indicate that it depends on both the acoustic model and pore size, as different loss mechanisms are involved. Both acoustic models predict variations with microstructure, indicating that the attenuation is underdetermined by clinical CT for both models. Related to possible clinical simulations, our results show that attenuation can differ by up to a factor of two between acoustic models at a single porosity, which exceeds the target precision of 11% to reduce focal pressure errors to 5% when using single-element transducers (Robertson *et al* 2017). This highlights the importance of using accurate acoustic models in estimating attenuation and the subsequent beam power in clinical simulations.

The two acoustic models produce different phase velocity-porosity relationships (Figure 4) due to different phase contributions related to differing scattering path lengths. The viscous model's phase velocity incorporates the average velocity along the path length, which lengthens with increased scattering. As such, the velocity decreases with increasing porosity and is slower for larger pore phantoms, which have larger scattering path lengths. In contrast, the viscoelastic model produces less variation and faster phase velocities at a given pore size and porosity. The reduced pore size dependence reflects the decreased role of scattering path length, as viscous and viscoelastic scattering differ, as described above. Furthermore, mode conversion may reduce the overall scattering path length, resulting in faster overall transmission; however, this is offset by the slower shear wave velocity. These results demonstrate that the phase velocity dependence on microstructure decreases with the viscoelastic models. Related to possible clinical simulations, our results show that the difference between viscous and viscoelastic phase velocities exceeds the target 4% uncertainty to reduce focal position error to 1.5 mm for a single-element transducer (Robertson *et al* 2017). This further emphasizes the need to utilize accurate acoustic models for phase velocity estimation.

We observed both positive and negative velocity dispersion, which depended, in part, on the acoustic model. The maximum dispersion rate observed in Figure 2 was 182 (m/s)/MHz, which produces minimal phase uncertainty over the typical frequency and skull thickness ranges for transcranial focusing. We note that dispersion is valuable in validating numerical methods and gaining insight into microstructural interactions (Haïat and Naili 2011, Waters and Hoffmeister 2005). For instance, negative dispersion is sometimes termed "anomalous" because it contradicts the prediction of logarithmic positive dispersion based on K-K relationships (Marutyan *et al* 2006).

Interestingly, the viscoelastic model's phase velocity result is more consistent with empirical relationships reported in the literature. Previous empirical studies of the phase velocity relationship to CT HUs have assumed linearity with density and porosity, based on the assumption of a fluid (Aubry *et al* 2003, Carter and Hayes 1977). Our study suggests this approximation may be appropriate, but the linearity and reduced sensitivity to microstructure may arise from viscoelastic interactions. These underlying mechanisms may explain why there is some consensus in the existing empirical phase velocity as determined by CT HUs (Leung *et al* 2019).

This study's methodology improved upon traditional through-transmission measurement approaches by utilizing the time-distance matrix for transmission. The time-distance matrix requires measuring the complex pressure amplitude within the bone, which can only be achieved using virtual sensors. However, it overcomes the challenge of removing incoherent and planar reflections to isolate the coherent ('ballistic') wave. Our previous study reported an insertion loss instead of an attenuation, in part because the resonant standing waves strongly affected the transmitted amplitude (Clinard *et al* 2025b). In this study, we improve on this prior result by reporting an attenuation in units of Np/cm. Further, our previous phase velocity estimates assumed the ballistic wave was stronger than incoherent scattering, an assumption that fails in larger pore phantoms that exhibit strong scattering (Clinard *et al* 2025a). This study's method more effectively measures phase velocity, as the ballistic wave is isolated at each distance using averaging and a time window. We note that our method still produces considerable variability within larger pore phantom fluid simulations, ranging from 20% to 50% porosity (Figure 4), which is attributed to an increased dependence on the spatial arrangement of the pores due to enhanced scattering, resonance, and possible wave-guide effects.

There are several limitations to our study. The digital phantoms employed simplified representations of skull bone, which allow their characterization by just two global parameters: porosity and pore size. This enables the inference of skull microstructure's effect on attenuation and phase velocity, which would not be possible in a more complicated heterogeneous phantom. This approach is well validated in prior literature (Larsson *et al* 2014). Also, we assign a uniform porosity to each phantom. As a result, we do not attempt to accurately model depth-dependent porosity (Alexander *et al* 2019) or variable pore size and anisotropy (Chen *et al* 2013). Because these phantoms don't fully represent real skull bone, the reported attenuation and phase velocity values should not be used to define a quantitative relationship between attenuation/velocity and porosity. Rather, they should be used to draw conclusions about how each simulation approach predicts acoustic transmission through heterogeneous skull bone. In addition, the simulation methods incorporate several modeling approximations. We used an attenuation power-law exponent of 2, which results in dispersionless material behavior under k-Wave's Kramers–Kronig implementation, to isolate microstructural effects on phase velocity dispersion.

Nonlinear effects were excluded from our simulations, as is appropriate for pressure amplitudes less than 1 MPa (Aubry *et al* 2022). Viscoelastic simulations have increased complexity and therefore a corresponding higher memory cost (additional elastic properties and stress-tensor propagation).

While viscous and viscoelastic k-Wave models yield different results, we cannot clearly determine which model is more accurate. Viscoelastic models include wave mode coupling at the bone-marrow interfaces, and as a result, may be more accurate (Haire and Langton 1999). Future work is needed to explore the accuracy of various viscous and viscoelastic models and their corresponding acoustic properties.

## Conclusion

This study demonstrated that viscous and viscoelastic models predict different relationships between bone microstructure and the velocity and attenuation of bone. Both models predict variations in attenuation and phase velocity due to microstructure that clinical CT cannot resolve; however, the viscoelastic model suggests that these variations may be minor when considering phase velocity. While we cannot yet determine which model is more accurate, it is clear that viscoelastic interactions affect the ultrasound propagation. By carefully considering the differences in acoustic models, future acoustic parameter estimation methods may better account for skull microstructure, thereby leading to safer and more effective transcranial treatment.

## Acknowledgements

Authors gratefully acknowledge funding from the Mark H. Huntsman Endowed Chair and NIH Grants F31DE032916, R21EB033117, and R01EB028316.

## Conflict of Interest Statement

The authors have no relevant conflicts of interest to disclose.

## Data Availability Statement

The manuscript data is available upon reasonable request to the corresponding author.